\documentclass[aip, jap, 10pt, twocolumn, reprint, superscriptaddress, amsmath, amssymb]{revtex4-2} 
\usepackage{graphicx}% Include figure files
\usepackage{dcolumn}% Align table columns on decimal point
\usepackage{bm}% bold math
\usepackage{soul}

\usepackage{float}
\usepackage{hyperref}% add hypertext capabilities
\usepackage[usenames]{xcolor}

\usepackage{multirow}
\usepackage{threeparttable}
\begin{document}
\newfloat{scheme}{hhtbp}{lof}
\floatname{scheme}{\small SCH.~}

\preprint{APS/123-QED}

\title{On the importance of electron-electron and electron-phonon scatterings and energy renormalizations during carrier relaxation in monolayer transition-metal dichalcogenides}

\author{J{\"o}rg Hader}
\email[Author to whom correspondence should be addressed: ]{jhader@acms.arizona.edu}
\affiliation{%
Wyant College of Optical Sciences, University of Arizona, 1630 E. University Blvd., Tucson, Arizona 85721, USA%\\This line break forced% with \\
}%

\author{Josefine Neuhaus}%
\affiliation{ 
Department of Physics and Material Sciences Center, Philipps-University Marburg, Renthof 5, 35032 Marburg, Germany%\\This line break forced with \textbackslash\textbackslash
}%

\author{Jerome V. Moloney}
\affiliation{%
Wyant College of Optical Sciences, University of Arizona, 1630 E. University Blvd., Tucson, Arizona 85721, USA%\\This line break forced% with \\
}%

\author{Stephan W. Koch}
\affiliation{ 
Department of Physics and Material Sciences Center, Philipps-University Marburg, Renthof 5, 35032 Marburg, Germany%\\This line break forced with \textbackslash\textbackslash
}%

\date{\today}%

\begin{abstract}
An $\it{ab \,\, initio}$ based fully microscopic many-body approach is used to study the carrier relaxation dynamics in monolayer transition-metal dichalcogenides. Bandstructures and wavefunctions as well as phonon energies and coupling matrix elements are calculated using density functional theory. The resulting dipole and Coulomb matrix elements are implemented in the Dirac-Bloch equations to calculate carrier-carrier and carrier-phonon scatterings throughout the whole Brillouin zone. It is shown that carrier scatterings lead to a relaxation into hot quasi-Fermi distributions on a single femtosecond timescale. Carrier cool down and inter-valley transitions are mediated by phonon scatterings on a picosecond timescale. Strong, density-dependent energy renormalizations are shown to be valley-dependent. For MoTe$_2$, MoSe$_2$ and MoS$_2$ the change of energies with occupation is found to be about 50$\%$ stronger in the $\Sigma$ and $\Lambda$ side valleys than in the $K$ and $K'$ valleys. However, for realistic carrier densities, the materials always maintain their direct bandgap at the $K$ points of the Brillouin zone.

\end{abstract}

%\pacs{71.15-m, 71.15.Mb, 71.20.Nr, 71.55.Eq}

\maketitle

%\tableofcontents

\section{Introduction}

Monolayer transition-metal dichalcogenides (TMDCs) have a variety of properties that make them very interesting for applications in opto-electronic devices. Two of the most important characteristics are their direct bandgap and the exceptionally strong Coulomb interaction due to inefficient screening outside the monolayer plane. The highly efficient optical coupling leads to a near-bandgap absorption of up to 10-20$\%$ from a single layer\cite{tenpcabsorption1,tenpcabsorption2} which makes the material attractive for applications like photo-detectors\cite{photodetector1,photodetector2,photodetector3} or solar cells\cite{solarcell1,solarcell2}. The prominent optical coupling also leads to strong luminescence and, consequently, promising performance as light-emitting diodes\cite{led1,led2,led3}. Monolayer TMDCs have also been shown to be able to provide strong optical gain\cite{gain1,gain2} which enables them to be used as active material in lasers\cite{laser1,laser2,laser3}. 

%%%%%%%%%%%%%%%%%%%%%%%%%%%%%%%%%%%%%%%%%%%%%%%%%%%%%%%%%%%%%
\begin{figure}[htp]
\begin{center}
\includegraphics[width=1.00\linewidth]{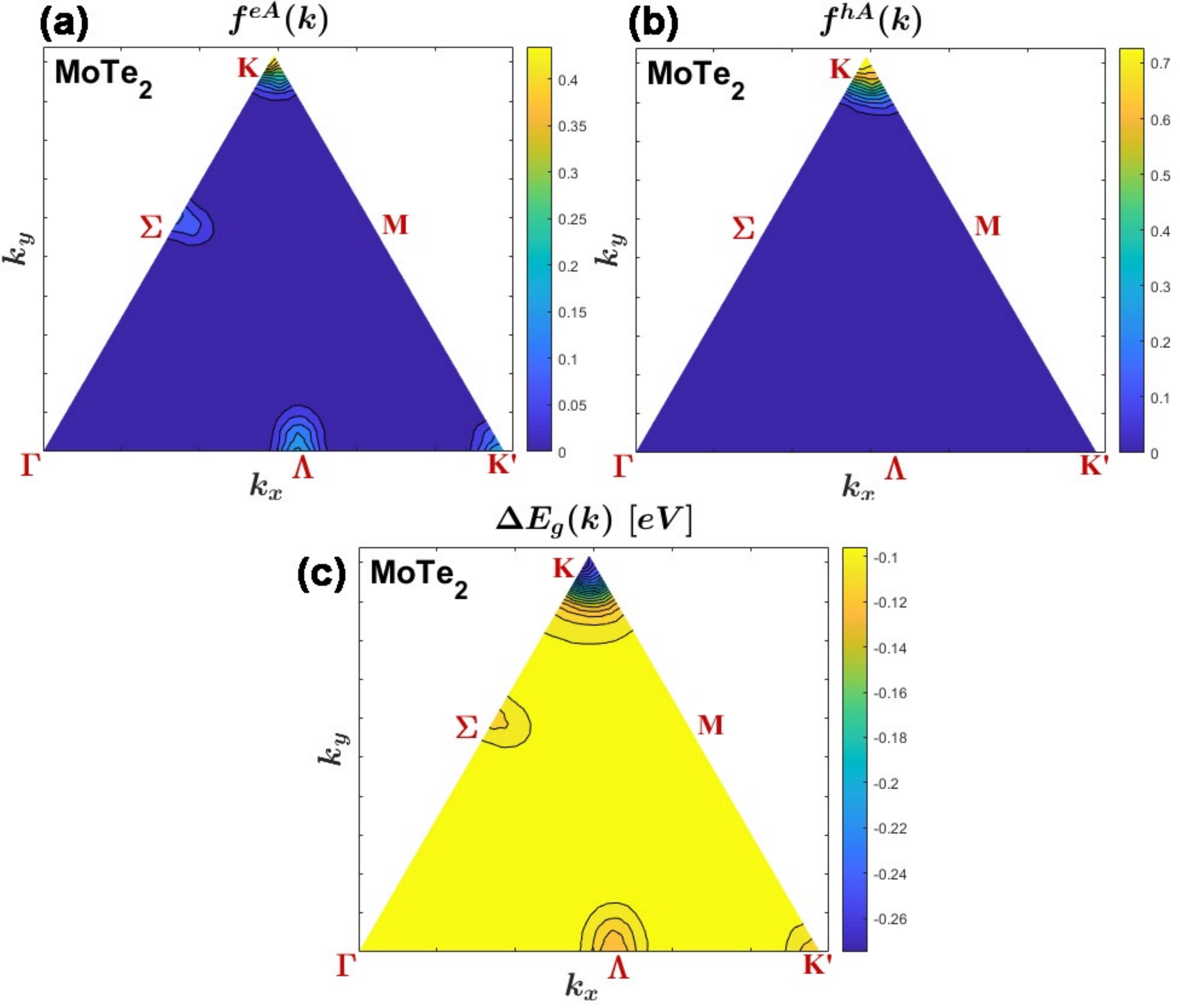}
\end{center}
\caption{(a)/(b): Electron/hole occupations of the A-band for a total electron/hole density of $12.5\times 10^{12}/$cm$^2$ in a monolayer MoTe$_2$. Carriers are assumed to be in thermal equilibrium Fermi distributions at 300$\,$K. (c): renormalization of the A-bandgap in the presence of the carriers.}
\label{fig1}
\end{figure}
%%%%%%%%%%%%%%%%%%%%%%%%%%%%%%%%%%%%%%%%%%%%%%%%%%%%%%%%%%%%%
The strong Coulomb interaction also enables the possibility to tune the bandgap over a wide spectral range. Injecting carriers into the system leads to energy renormalizations on the order of the exciton binding energy which in these materials is of the order of hundreds of meV\cite{gapren1,gapren2,gapren3,gapren4}. Fig.\ref{fig1} gives an example for the energy renormalizations in a monolayer of MoTe$_2$. Here, equal electron and hole densities of $12.5\times 10^{12}/$cm$^2$ were placed in Fermi dstributions at 300$\,$K. Figs.\ref{fig1} (a) and (b) show the carrier distributions in the band that is energetically closest to the bandgap at the $K$-point (A-band). Mirroring these results along the $\Gamma$-$M$ line gives the results for the B-band which has the opposite spin and the same gap at $K'$ as the A-band at $K$.  As can be seen from Fig.\ref{fig1} (c), the presence of the quasi-equilibrium carriers reduces the bandgap by about 100$\,$meV throughout the whole Brillouin zone (BZ). The renormalization is strongly enhanced in the vicinity of high carrier occupations, reaching over 260$\,$meV at the $K$-point. 

The strong energy renormalizations open the interesting possibility to tune the operating wavelength through controlled carrier injection. Also, the energetic order of the valleys could potentially be changed, e.g., by exciting resonantly with a specific valley transition and/or using polarized, spin-valley sensitive excitation\cite{valley1,valley2}. However, valley specific renormalizations require non-thermal carrier distributions. Once pump injected carriers have thermalized into Fermi distributions with a global Fermi level throughout the BZ, the energies of individual valleys will only be determined by the total carrier density. A potential drawback of the density dependent renormalizations was pointed out in Refs.\cite{gain1,dirindir1}. If the energies of a side valley of the electron A-band renormalize faster with increasing carrier density than the energy at the direct bandgap at $K$, it could potentially happen that the side valley is lowered below the $K$-valley and the material becomes optically indirect. In thi case, the pump injected electrons would predominantly occupy the side valley, and, since the hole bandstructure does not have side valleys, electrons in them are not available to recombine optically. This will degrade the material's performance for applications like light emitting diodes or lasers. As we will show here, whether such a transition will occur strongly depends on details of the bandstructure.

Valley- and excitation-dependent renormalizations for carriers in static thermal equilibrium distributions were discussed in Ref.\cite{dirindir1} using similar first principles based microscopic models as we employ here. In Ref.\cite{ref4} a similar model was extended in order to be able to study the nonequilibrium carrier relaxation due to electron-electron scatterings. However, inter-valley carrier transitions involve large momentum transfer which is predominantly mediated by scatterings involving optical and acoustical phonons. These were not included in Ref.\cite{ref4}. The influence of phonon-assisted inter- and intra-valley scatterings were investigated in Refs.\cite{ref5,ref6,ref7}, but only for carriers in equilibrium distributions. In Ref.\cite{gain2} electron-electron and electron-phonon scatterings were included to study nonequilibrium carrier relaxation. However, that investigation was limited to intra-valley dynamics using a simplified one-dimensional bandstructure model.

Here, we employ a fully microscopic many-body model that includes electron-electron and electron-phonon scatterings throughout the full BZ. The model is  based on input from first principle density functional theory (DFT) for bandstructures, electron wavefunctions and phonon energies and coupling matrix elements. The model is used to determine timescales for the relaxation of carriers that are excited above the bandgap. It also yields times for inter-valley and intra-valley scatterings that lead to global quasi equilibrium distributions as well as for the subsequent cool-down to the ambient temperature by removal of excess energy via phonon emission.

Details of the model are outlined in Sec.\ref{sec_theo} where we will also show results of the DFT calculations. Results of the many-body model are discussed in Sec.\ref{sec_results} for monolayer MoTe$_2$, MoSe$_2$ and MoS$_2$. In Sec.\ref{sec_artexc} we present numerical results assuming an artificial static carrier distribution to initialize the system in order to be able to clearly distinguish between the excitation related dynamics and the subsequent carrier relaxation. Carrier scattering processes are turned on after the initialization. In Sec.\ref{sec_artaboexc} we examine the case of an initial distribution that is energetically located above the barriers separating different bandstructure valleys. We determine the characteristic carrier relaxation times and show the influence of electron-electron scatterings versus electron-phonon scatterings due to optical and acoustical phonons at various stages of the relaxation process. In Sec.\ref{sec_artresexc} we present results where we assume excitation at the bandgap in the K-valley and investigate the subsequent carrier transfer to the K', $\Sigma$ and $\Lambda$ valleys. After that, in Sec.\ref{sec_optexc}, we simulate the situation where the system is excited dynamically using a 50$\,$fs optical pulse which allows us to identify characteristic details of the relaxation dynamics might be observable in the experiment. Finally, in Sec.\ref{sec_eneren}, we investigate the energy renormalizations after full carrier thermalisation for various excitation levels in order to see if a transition from direct to indirect bandgap occurs. We summarize our results in Sec.\ref{sec_summary}.

\section{Theoretical models}
\label{sec_theo}

The theoretical approach used here is based on the Dirac-Bloch equations (DBE) as described in Ref. \cite{gain2} and references therein. The DBE are the equations of motions for the microscopic polarizations, $p_{\textbf{k}i}$, and the occupation probabilities for electrons/holes, $f^{e/h}_{\textbf{k}i}$:
\begin{eqnarray}
\label{dbe}
i\hslash\frac{d}{d t} p_{\textbf{k}i} &=& \left(\varepsilon^e_{\textbf{k}i}-\varepsilon^h_{\textbf{k}i}\right)p_{\textbf{k}i} - \left( 1-f^{e}_{\textbf{k}i}-f^{h}_{\textbf{k}i}\right)\Omega_{\textbf{k}i} \nonumber \\
 &&{ +} i\hslash\left.\frac{d}{d t} p_{\textbf{k}i}\right|_\mathrm{corr}\,,\\
\frac{d}{d t} f^{e/h}_{\textbf{k}i} &=& -\frac{2}{\hslash}\,\text{Im}\left(\Omega_{\textbf{k}i} p^*_{\textbf{k}i}\right) { +} \left.\frac{d}{d t}f^{e/h}_{\textbf{k}i}\right|_\mathrm{corr}.
\end{eqnarray}
Here, $\Omega$ are the renormalized Rabi frequencies
\begin{eqnarray}
\label{renf}
 \Omega_{{\bf k}i} &=& \frac{e}{m_0c}{\bf A\cdot  {\boldsymbol{\mu}}_{{\bf k}i}}-\sum\limits_{\bf k'} V^{ehhh}_{\bf k-k';k';k}(f^e_{{\bf k'}i}-f^h_{{\bf k'}i})\nonumber \\
  && -\sum\limits_{\bf k'}\left[ V^{ehhe}_{\bf k-k';k';k}p_{{\bf k'}i}+ V^{eehh}_{\bf k-k';k';k}p^*_{{\bf k'}i}\right],
\end{eqnarray}
where ${\boldsymbol{\mu}}_{{\bf k}i}$ are the dipole matrix elements between the electron and hole bands with band index $i$ and momentum ${\bf k}$, and $\varepsilon$ are the renormalized energies:
\begin{eqnarray}
\label{rene}
\varepsilon^{e/h}_{\textbf{k}i} &=& \epsilon^{e/h}_{\textbf{k}i} - \sum_{\textbf{k}'} \left[V^{eeee/hhhh}_{\bf k-k';k';k}-V^{eheh/hehe}_{\bf k-k';k';k}\right]f^{e/h}_{{\bf k'}i} \nonumber \\
 &&+ \sum_{\textbf{k}'} \left[V^{ehee/hhhe}_{\bf k-k';k';k}p_{{\bf k'}i}+c.c.\right].
\end{eqnarray}
Coulomb matrix elements of the type $V^{ehhh}$ in Eq.(\ref{renf}) and $V^{ehee/hhhe}$ in Eq.(\ref{rene}) mediate Auger-like processes. Coulomb terms of the type $V^{eehh}$ in Eq.(\ref{renf}) and $V^{eheh/hehe}$ in Eq. (\ref{rene}) represent pair creation and annihilation processes. Whereas these processes can be neglected in systems with weaker Coulomb interaction, like III-V semiconductors, they are important in TMDCs. Here, they lead to a non-trivial ground state deviating from the case of zero occupations and polarizations. The dominant influence of these terms is to contribute to a renormalisation of the  hole bands which we take into account by calculating the so-called Coulomb hole. We have tested that these Auger- and pair-processes have a negligible impact on the carrier dynamics and can be neglected there.
 
 The term involving $V^{ehhe}$ in Eq. (\ref{renf}) represents the renormalisation of the optical coupling as known from the classical semiconductor Bloch equations (SBE). It leads to the generation of bound excitonic states below the bandgap and to the Coulomb-enhancement of the continuum absorption. As in the SBE, the term involving $V^{eeee/hhhh}$ in Eq. (\ref{rene}) leads to  the exitation dependent energy renormalisation that we will investigate in detail here.

The terms marked $|_{corr}$ summarize many-body correlations beyond the Hartree-Fock level. They contain carrier scatterings due to electron-electron and electron-phonon interactions. The higher order correlations are also responsible for the { plasma-}screening of the Coulomb interaction. In monolayer TMDCs, the correct inclusion of these higher order terms is of particular importance due to the exceptional strength of the Coulomb interaction.  We treat these terms and the resulting screening as outlined in Ref.\cite{gain2}.

For the calculations as shown here, scatterings are only taken into account explicitly for the carrier distributions. Polarizations are only included for the optical excitation of carriers. Since our focus is on the subsequent incoherent dynamics of the distributions and not details of the excitation, we approximate the scatterings that lead to the dephasing of the polarisations by a simple dephasing rate.

The electron-electron scattering equations solved here are outlined in Ref.\cite{gain2}. Whereas in that reference only one branch of dispersionless optical phonons  was taken into account for the electron-phonon scattering, here, we include all optical and acoustical phonon branches. The generalized electron-phonon scattering equations take the form:
\begin{eqnarray}
\left. \frac{d}{dt} f_{{\bf k}i}\right\vert_{\mathrm{corr.}}^{\mathrm{ph.}}&=&
	 \frac{2\pi}{\hbar}\sum_{{\bf q},m}g^{i,m}_{{\bf k; q}}\tilde{g}^{i,m}_{{\bf k; q; k+q}}
	\mathcal{D}\left(\tilde{\varepsilon}_{{\bf k+q}i}-\tilde{\varepsilon}_{{\bf k}i}-\hbar\omega^m_{{\bf q}}\right)\nonumber \\
	&& \times \left[(n^m_{{\bf q}}+1)f_{\mathbf{k+q}i}\bar{f}_{\mathbf{k}i}-
	n^m_{{\bf q}}f_{\mathbf{k}i}\bar{f}_{\mathbf{k+q}i}\right] \nonumber \\
&+&
	\frac{2\pi}{\hbar}\sum_{{\bf q},m}g^{i,m}_{{\bf k; q}}\tilde{g}^{i,m}_{{\bf k; q; k}}
	\mathcal{D}\left(\tilde{\varepsilon}_{{\bf k-q}i}-\tilde{\varepsilon}_{{\bf k}i}-\hbar\omega^m_{{\bf q}}\right)\nonumber \\
	&& \times \left[n^m_{{\bf q}}f_{\mathbf{k-q}i}\bar{f}_{\mathbf{k}i}-
	(1+n^m_{{\bf q}})f_{\mathbf{k}i}\bar{f}_{\mathbf{k-q}i}\right],
\label{eq:phsc}
\end{eqnarray}
where we use the abbreviation $\bar{f}$ for $(1-f)$ and combined the band indices $i$ with the electron/hole index $e/h$. $\pi\mathcal{D}(x)=\frac{\eta}{x^2+\eta^2}$ denotes the numerical energy-conserving function, where we include a phenomenological broadening of $\eta=50\,$meV. Numerical tests showed that the exact value of this broadening was insignificant for the final results. The renormalized transition energies $\tilde{\varepsilon}$ are calculated as in Eq.(\ref{rene}) but with Coulomb matrix elements which include plasma-screening due to excited carriers in addition to the screening from the dielectric environment which is already included in the matrix elements $V$. The index $m$ labels the nine phonon branches{ \cite{phonondisp1,phonondisp2}}, all of which are taken into account in our numerical evaluations. The phonon energies are denoted by $\hbar\omega^m_{{\bf q}}$. $n^m_{\mathbf{q}}$ are the phonon occupation numbers. The quantities $g$ and $\tilde{g}$ are the unscreened and screened phonon coupling matrix elements, respectively. The additional ${\bf k}$-index on the phonon coupling matrix elements arises from the fact that the momentum dependence of the electron wavefunctions is included explicitly here while the studies in Ref.\cite{gain2} exclusively focused on the $K$-valley and used the electron wavefunctions at this point only.

{ In Eq.(\ref{eq:phsc}), the first term of the second line corresponds to scattering of carriers into the state ${\bf k},i$ via phonon emission and the second term represents out-scattering via phonon absorption. The first term of the fourth line represents in-scattering via phonon absorption and the last term out-scattering via phonon emission. For all processes the initial state has to be occupied, i.e. the occupation $f$ has to be non-zero, and the final state has to be at least partially empty, $(1-f)>0$. Energy has to be conserved and the scattering probability is given by the phonon coupling matrix elements $g$.}

The unrenormalized single-particle energies, $\epsilon$, are calculated via DFT using the Vienna {\it ab initio} simulation package (VASP)\cite{vasp1,vasp2,vasp3,vasp4,vasp5}. Details of these calculations are described in Ref.\cite{gain2} for MoTe$_2$ and executed analogously for the other materials. Besides the bandstructures, we also extract the dipole matrix elements and the wavefunctions needed to evaluate the Coulomb matrix elements from these calculations. { The full k-dependence throughout the BZ is taken into account in all calculations for the single particle energies and dipole matrix elements. Unlike in Ref.\cite{gain2}, no simplification in terms of a massive Dirac fermion model with fitted bands is used. For the Coulomb matrix elements we evaluate the wavefunction-dependent form factors only in the vicinity of the $K$ and $\Sigma$ point and assume the same factors for the $K'$ and $\Lambda$ valleys. In regions away from these points the form factor of the nearest valley is used.}

%%%%%%%%%%%%%%%%%%%%%%%%%%%%%%%%%%%%%%%%%%%%%%%%%%%%%%%%%%%%%
\begin{figure}[htp]
\begin{center}
\includegraphics[width=1.00\linewidth]{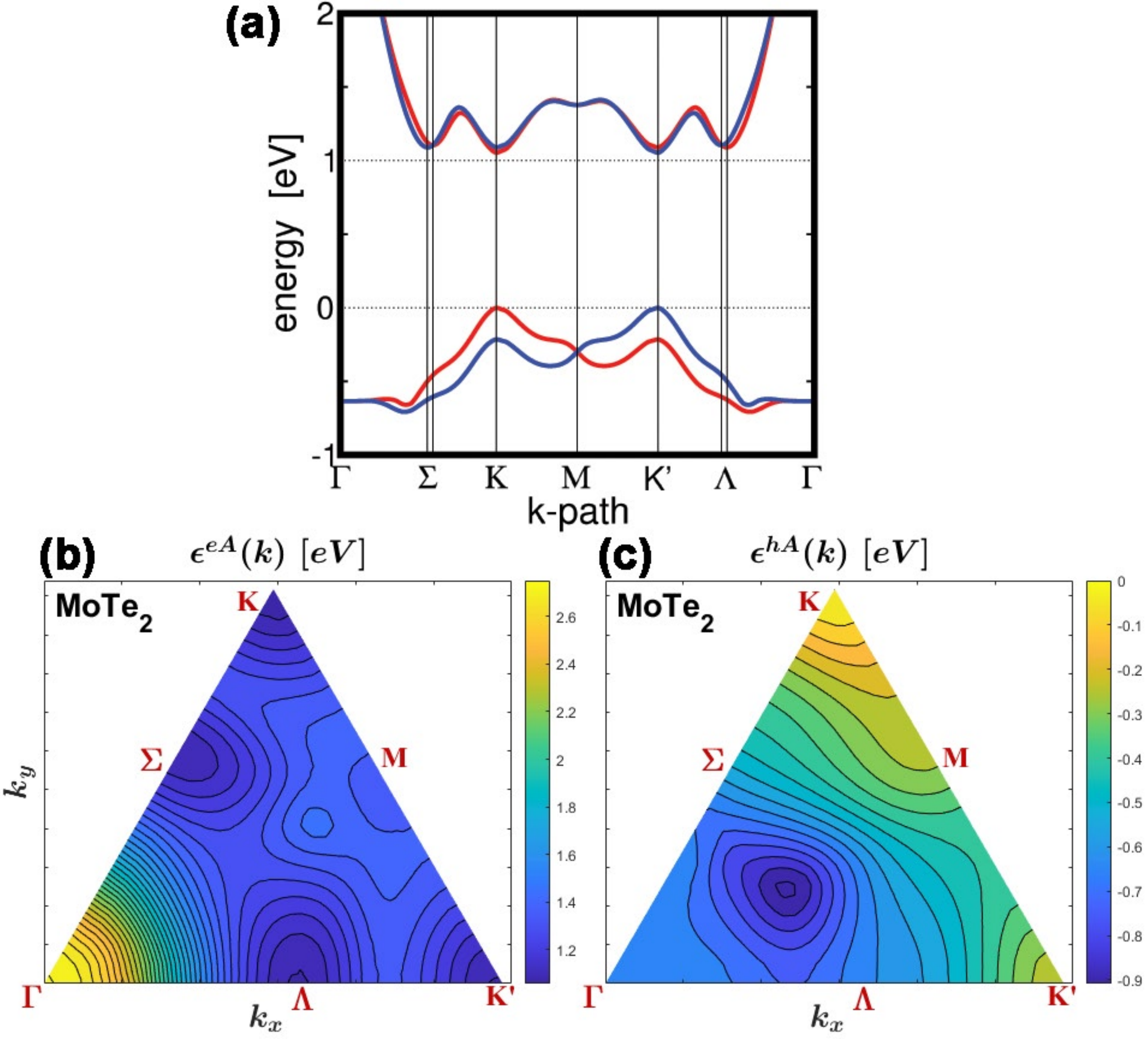}
\end{center}
\caption{{ (a): lowest two electron and highest two hole bands in a monolayer of MoTe$_2$ along a one-dimensional path through the BZ. Red/blue: bands for carriers with spin up/down (A/B-band). Bottom: energy of the electron (b) and hole (c) A-bands in the irreducible sector of the BZ.} Spacing between contour lines is 50$\,$meV.}
\label{fig2}
\end{figure}
%%%%%%%%%%%%%%%%%%%%%%%%%%%%%%%%%%%%%%%%%%%%%%%%%%%%%%%%%%%%%

 For all materials investigated here, the bandstructure has the same basic features. As representative example, Fig.2 shows the calculated unrenormalized electron and hole bands for the case of a monolayer of MoTe$_2$. The lowest two electron bands and highest two hole bands are related through time reversal symmetry. Spin-orbit splitting lifts the degeneracy of the bands with opposite spin at $K$ and $K'$. The electron band with the lowest energy at $K$ and the hole band with the highest energy at $K$ are referred to as the A-bands. Carriers in the so-called B-band have the opposite spin from those in the A-band. Energies of the B-band can be derived from those of the A-band by mirroring perpendicular to the $\Gamma-$M line. For excitation with linear polarized light the the physics within the A-band at $K$ is the same as that within the B-band at $K'$. Thus, we reduce our presentations to the the A-band in the following.

There are three additional minima in the electron A-band structure at $K'$, $\Sigma$ and $\Lambda$.  The hole A-band has maxima at the $K$- and $K'$-points and, additionally, a local maximum at the $\Gamma$-point. The energies of the unrenormalized electron A-bands at critical points of the BZ are listed for all materials investigated here in Table \ref{tabenee}. Hole energies in the A-band are listed in Table \ref{tabeneh}.
\begin{table}
\caption{\label{tabenee}Calculated unrenormalized energies in the electron A-band at critical points of the BZ and at energy maxima, $m_{P;P'}$, between points $P$ and $P'$. Energies in [eV].}
\begin{tabular}{l|ccccccc}
 & $K$ & $K'$ & $\Lambda$ & $\Sigma$ & $m_{K;\Sigma}$ & $m_{K';\Lambda}$ & $m_{K;K'}$ \\
\hline
MoTe$_2$ & 1.058 & 1.093 & 1.091 & 1.108 & 1.323 & 1.363 & 1.410 \\
MoSe$_2$ & 1.446 & 1.468 & 1.496 & 1.474 & 1.769 & 1.784 & 1.900 \\
MoS$_2$ & 1.708 & 1.711 & 1.821 & 1.892 & 2.156 & 2.160 & 2.309
\end{tabular}
\end{table}
\begin{table}
\caption{\label{tabeneh} Calculated unrenormalized energies in the hole A-band at $K$, $K'$ and $\Gamma$ and at energy minima, $m_{P;P'}$, between points $P$ and $P'$. Energies in [eV].}
\begin{tabular}{l|cccccc}
 & $K$ & $K'$ & $\Gamma$ & $m_{K;K'}$ & $m_{K;\Gamma}$ & $m_{K';\Gamma}$ \\
\hline
MoTe$_2$ & 0.000  & -0.214 & -0.393 & -0.633 & -0.656 & -0.705 \\
MoSe$_2$ & 0.000 & -0.183 & -0.670 & -0.435 & -0.777 & -0.853 \\
MoS$_2$ & 0.000  & -0.146 & -0.153 & -0.706 & -0.838 & -0.919
\end{tabular}
\end{table}

It should be noted that the exact valley energies are very sensitive to details of the DFT calculation. In turn, the excitation dependent energy renormalizations that will be studied in Sec.\ref{sec_eneren} are strongly influenced by the exact separations between main and side valleys. One major aspect concerning the offset between the different valleys is the lattice constant. Already differences of one percent in the lattice constant change the calculated DFT bandgap in the range of 100$\,$meV and may result in the transition from a direct to an indirect semiconductor.\cite{strain1,strain2,strain3}. In order to be as precise as possible, we included van-der-Waals interaction via Grimme's dispersion correction, resulting in relaxed in-plane lattice constants of 3.50\AA$\,$ for MoTe$_2$, 3.28\AA$\,$ for MoSe$_2$, and 3.15\AA$\,$ for MoS$_2$, which is comparable to experimental findings.\cite{masses} Based on these parameters, we find a direct bandgap at $K$ for all materials investigated here and a separation from the side valleys at $\Sigma$ and $\Lambda$ that exceeds the room temperature thermal energy of about 26$\,$meV.

Phonon energies and coupling matrix elements are calculated using the DFT software Quantum Espresso\cite{qe1,qe2}. For the non-collinear Kohn-Sham wavefunctions with spin-orbit coupling, the plane-wave basis set with a 49 Rydberg energy cut-off, Perdew-Burke-Ernzerhof type generalized gradient approximated exchange-correlation functional, and projector augmented wave (PAW) method with full-relativistic potentials were used. The resulting phonon dispersions were found in good agreement with the literature (see e.g. Refs.\cite{phonondisp1,phonondisp2}). { The full ${\bf k}$- and ${\bf q}$-dependence of the phonon dispersions and coupling matrix elements are taken into account in all calculations.}

%The calculated phonon dispersions for the example of a monolayer of MoTe$_2$ are shown in Fig.\ref{figphondisp}. They are in very good agreement with the results of Ref.\cite{phonondisp}.
%%%%%%%%%%%%%%%%%%%%%%%%%%%%%%%%%%%%%%%%%%%%%%%%%%%%%%%%%%%%%
%\begin{figure}[htp]
%\begin{center}
%\includegraphics[width=1.00\linewidth]{mote2_phonon_disp}
%\end{center}
%\caption{Phonon dispersions for a monolayer MoTe$_2$.}
%\label{figphondisp}
%\end{figure}
%%%%%%%%%%%%%%%%%%%%%%%%%%%%%%%%%%%%%%%%%%%%%%%%%%%%%%%%%%%%%

\section{Results}
\label{sec_results}

In the following we examine results for monolayers of MoTe$_2$, MoSe$_2$ and MoS$_2$ suspended on a SiO$_2$ substrate and at a temperature of 300$\,$K. In all studies of the dynamics only the lowest two electron and highest two hole bands -- A and B -- are included. Excitation with linear polarized light is assumed which creates equal amounts of carriers at the $K$ and $K'$ points in bands of opposite spin.

\subsection{Artificial Excitation}
\label{sec_artexc}

Here, we initialize the carrier distributions with an artificial, static distribution that allows to clearly determine timescales in the subsequent  relaxation as well as the importance of various scattering processes within it. The initial distributions are given by the formula:
\begin{equation}
\label{eq_inif}
f_{{\bf k}i} = \frac{f_0\mu^2_{{\bf k}i}}{\mu^2_{max}} \exp\left(-\frac{1}{2}\left(\frac{\epsilon^{e}_{{\bf k}i}-\epsilon^{h}_{{\bf k}i}-\hbar\omega_L}{\Delta}\right)^2\right),
\end{equation}
where  ${\bf k}$ is the two-dimensional momentum vector, $i$ is the band index and $\epsilon^{e/h}$ are the unrenormalized electron/hole energies.
These distributions approximate a case where carriers are optically excited at an energy $\hbar\omega_L$ and form a Gaussian distribution of width $\Delta$. The occupation probabilities are weighted with the square of the dipole matrix element $\mu_{{\bf k}i}$ between the $i$'th electron and hole band assuming a vanishing optical coupling between states with unequal spin. $\mu_{max}$ is the largest dipole matrix element between any states. For all cases studied here, a broadening $\Delta=66\,$meV is used.

\subsubsection{Above Bandgap Excitation}
\label{sec_artaboexc}

%%%%%%%%%%%%%%%%%%%%%%%%%%%%%%%%%%%%%%%%%%%%%%%%%%%%%%%%%%%%%
\begin{figure}[htp]
\begin{center}
\includegraphics[width=1.00\linewidth]{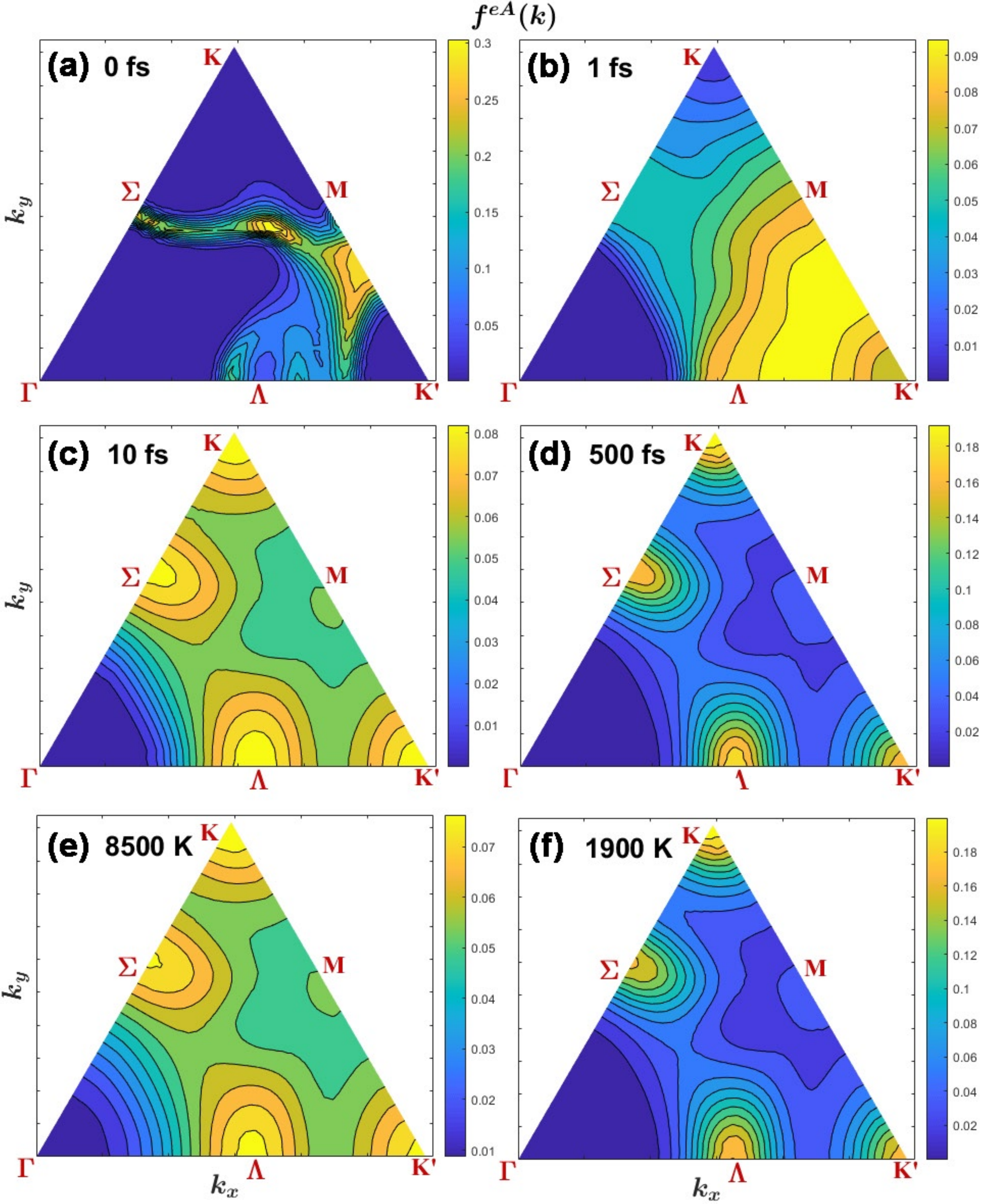}
\end{center}
\caption{(a), (b), (c), (d): Occupations of the electron A-band in a monolayer MoTe$_2$ at various times after initialisation according to Eq.(\ref{eq_inif}) with $f_0=1.28$. Labels give the time after the start of relaxation. (e), (f): Occupation of the electron A-band for the same total electron density as in (a)-(d), but using Fermi distributions with carrier temperatures of 8500$\,$K and 1900$\,$K, respectively.}
\label{fig3}
\end{figure}
%%%%%%%%%%%%%%%%%%%%%%%%%%%%%%%%%%%%%%%%%%%%%%%%%%%%%%%%%%%%%
Fig.\ref{fig3} (a) shows the distributions created in the electron A-band using Eq.(\ref{eq_inif}) for an excitation energy $\hbar\omega_L$ 800$\,$meV above the unrenormalized bandgap. Here, a scaling $f_0=1.28$ was used which leads to a total electron sheet carrier density of about $1 \times 10^{14}/$cm$^2$. Figs.\ref{fig3} (b), (c), and (d) show the distributions after 1$\,$fs, 10$\,$fs, and 500$\,$fs of relaxation, respectively. As can be seen in Fig. \ref{fig3} (b), carrier scatterings broaden the distribution dramatically on a single femtosecond timescale. The excitation high above the bandgap is ideal for fast relaxation. Initial states are highly occupied and final states at lower energies are mostly empty which eliminates slow down of the relaxation due to phase space filling in this initial phase. Also, carriers located in rather narrow momentum regions at high energies screen the Coulomb interaction far less efficiently than if the same amount of carriers are relaxed  and distributed throughout wide regions of the BZ. 

After only 10$\,$ fs the carriers have relaxed to the bottom of the band and assume Fermi-like distributions with maximum occupation at minimum energy (see Fig.\ref{fig3} (c)). The excess energy from the excitation leads to very hot distributions. Using Fermi distributions for the same electron density we find a very good match to the distribution after 10$\,$fs assuming a carrier temperature of 8,500$\,$K (see Fig.\ref{fig3} (e)). The carriers subsequently cool down due to phonon emission. As can be seen from Figs.\ref{fig3} (d) and (f), after 500$\,$fs the distribution can be matched assuming a carrier temperature of 1,900$\,$K. Subsequently, the carriers cool down further toward the lattice temperature of 300$\,$K on a picosecond timescale. 

%%%%%%%%%%%%%%%%%%%%%%%%%%%%%%%%%%%%%%%%%%%%%%%%%%%%%%%%%%%%%
\begin{figure}[htp]
\begin{center}
\includegraphics[width=1.00\linewidth]{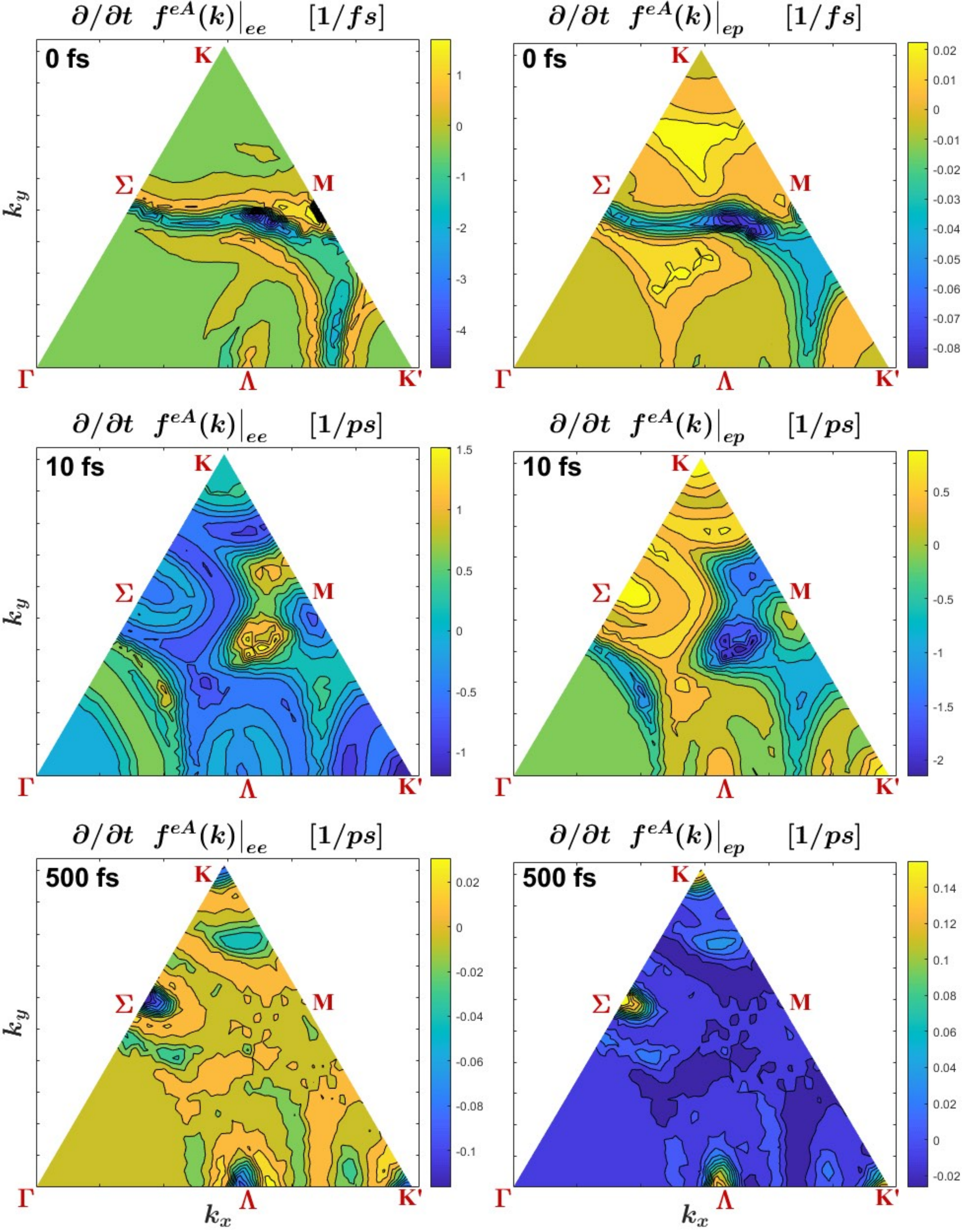}
\end{center}
\caption{Change of occupations of the electron A-band in a monolayer MoTe$_2$ at various times after initialisation as for Fig.\ref{fig3}. Left: change due to electron-electron scattering. Right: change due to electron-phonon scattering. Units for the top two (bottom four) panels are $[1/$fs$]$ ($[1/$ps$]$).}
\label{fig4}
\end{figure}
%%%%%%%%%%%%%%%%%%%%%%%%%%%%%%%%%%%%%%%%%%%%%%%%%%%%%%%%%%%%%
Fig.\ref{fig4} shows the individual contributions of electron-electron scatterings and of electron-phonon scatterings to the carrier relaxation at the start of the relaxation and after 10 and 500$\,$fs of relaxation. Initially, the relaxation is dominated by electron-electron scattering which broadens the localized distributions according to a  hot plasma temperature on a single femtosecond timescale. During this initial phase, the dynamic due to electron-phonon scatterings is about two orders of magnitude slower and, thus, irrelevant at that time. Once the carriers have relaxed into hot quasi-Fermi distributions after about 10$\,$fs, the overall dynamic slows down by two to three orders of magnitude (see the change of units in Fig.\ref{fig4} from $[1/$fs$]$ to $[1/$ps$]$). Here, electron-electron scattering and electron-phonon scattering become equally important.  The scatterings slow further down by roughly another order of magnitude once the distributions cool down near room temperature after a few picoseconds. In- and out-scatterings due to electron-electron and electron-phonon scatterings lead eventually to the detailed balance that defines equilibrium Fermi distributions.

\begin{figure}[htbp]
    \includegraphics[width=0.99\linewidth]{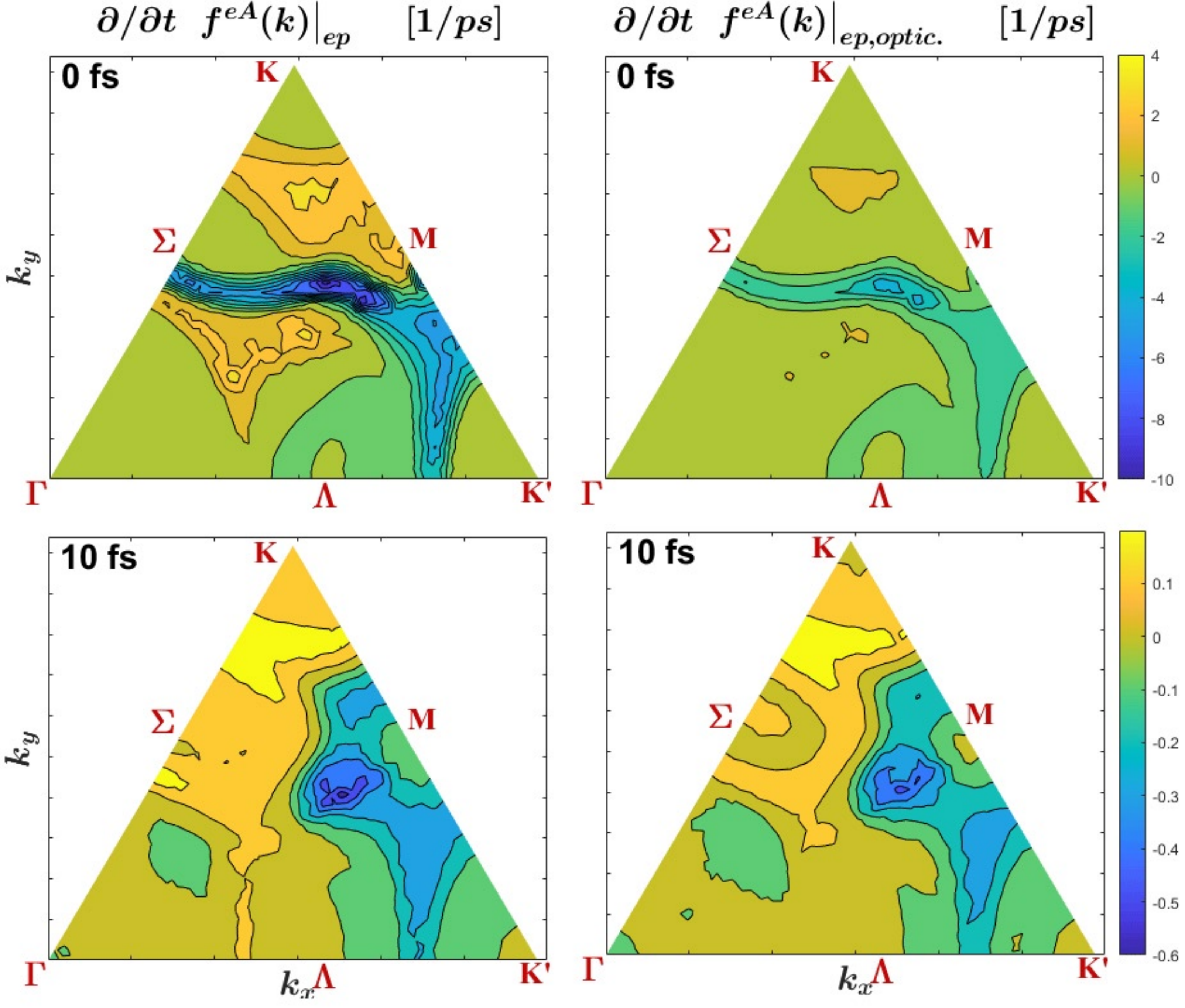}
    \caption{Change of occupations of the electron A-band in a monolayer MoTe$_2$ due to phonon scatterings only. Top: at the beginning of carrier relaxation.Bottom: after 10$\,$fs of relaxation. Initialisation as for Fig.\ref{fig3} but with $f_0=0.16$. Left: change due to all electron-phonon scatterings. Right: change only due to optical phonons.}
    \label{fig5}
\end{figure}

As can be seen in Fig.\ref{fig5}, during the first few femtoseconds of relaxation the phonon scatterings involving acoustic phonons are of similar importance as scatterings on optical phonons.  Here, the phonon scattering rates including all phonon contributions are about twice as high as when acoustical phonons are omitted. For the broadening of the initial distribution scatterings involving large momentum transfers are of particular importance. While acoustical phonons have rather limited coupling strength for small momentum transfers as they are typically involved in intra-valley relaxation, their coupling strength and energy increases with increasing momentum transfer which increase their importance for inter-valley scatterings. Without the acoustic phonons there are only very few scatterings far from the initial distribution, like into the area between $\Sigma$, $K$, and $M$ or between $\Sigma$, $\Lambda$ and $\Gamma$. After the carriers have assumed a hot quasi-Fermi distribution after about 10$\,$fs, the carrier relaxation becomes dominated by intra-valley scatterings. Here, scatterings involving optical phonons dominate the total electron-phonon interaction and the total phonon rates are almost identical to those in the absence of processes involving acoustical phonons.

For the comparison of phonon scattering contributions in Fig.\ref{fig5} we assumed an 8-times weaker excitation than for the results in Figs.\ref{fig3} and \ref{fig4}. Electron-electron scattering roughly scales with the density squared, while electron-phonon scattering scales about linearly with it. Thus, for this lower excitation, phonon scatterings are overall more relevant than in the earlier study. Also, it has been observed that for strong excitations the limited phonon density of states in the two-dimensional system leads to a hot phonon bottleneck\cite{hotphon1,hotphon2,hotphon3,hotphon4}. Here, a build-up of a nonequilibrium distribution for optical phonons can strongly limit their cooling power which, in turn, increases the relative importance of acoustical phonons. This hot phonon effect is not included in our current model.

%%%%%%%%%%%%%%%%%%%%%%%%%%%%%%%%%%%%%%%%%%%%%%%%%%%%%%%%%%%%%
%%%%%%%%%%%%%%%%%%%%%%%%%%%%%%%%%%%%%%%%%%%%%%%%%%%%%%%%%%%%%

\subsubsection{Resonant Excitation}
\label{sec_artresexc}

\begin{figure}[htbp]
    \includegraphics[width=0.99\linewidth]{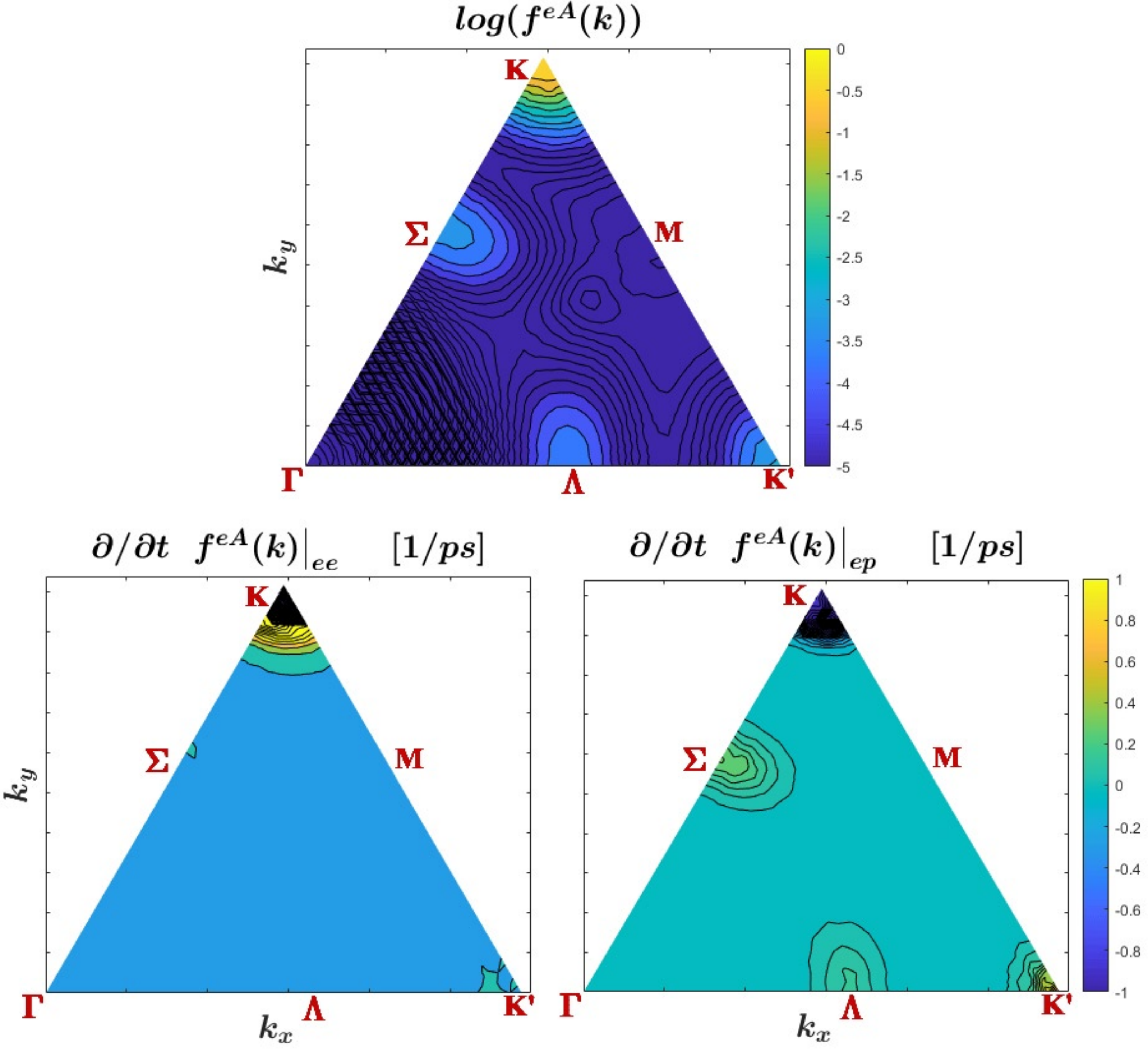}
    \caption{Top: logarithm of the occupation of the electron A-band in a monolayer MoTe$_2$ 2$\,$fs after initialisation with Eq.(\ref{eq_inif}) resonantly at the bandgap with $f_0=1.28$. Bottom left: change in occupations within the electron A-band due to electron-electron scattering. Bottom right: change due to electron-phonon scattering. Grid-line spacing is 0.05$/$ps in the lower two figures. { Please note that the solid black areas near $K$ in the lower two graphs are due to steep slopes resulting in overlapping contour lines and not data exceeding the color bar scale.}}
    \label{fig6}
\end{figure}

In order to study the inter-valley carrier transfer we place an artificial initial distribution almost exclusively in the $K$-valley of the A-band (and $K'$-valley of the B-band) and then calculate the scattering related carrier generation in the other valleys.
Fig.\ref{fig6} shows the occupations of the electron A-band in MoTe$_2$ 2$\,$fs after initialisation with Eq.\ref{eq_inif} resonantly at the bandgap, i.e.  with $\hbar\omega_L=(\epsilon^e_{K,A}-\epsilon^h_{K,A})$, and $f_0=1.28$.  This initialisation creates an occupation of about $0.75$ at $K$. After the initial 2$\,$fs of dynamics, scatterings have created electron occupations of about 0.0014 at $K$, 0.0003 at $\Sigma$ and 0.0007 at $\Lambda$.

Electron-electron scattering leads to a very fast initial relaxation within the $K$-valley to establish a Fermi-like distribution there on a  femtosecond timescale. However, it does not lead to significant inter-valley scattering due to the large involved momentum transfer. The rate is only about $0.1/$ps for electron-electron scattering from $K$ to $K'$  and even less for transitions to $\Sigma$ and $\Lambda$. Inter-valley transfer is dominated by electron-phonon scattering which during this initial phase creates carrier at a rate of about $0.8/$ps in the $K'$-valley, $0.2/$ps in the $\Lambda$-valley, and $0.3/$ps in the $\Sigma$-valley.
%%%%%%%%%%%%%%%%%%%%%%%%%%%%%%%%%%%%%%%%%%%%%%%%%%%%%%%%%%%%%
%%%%%%%%%%%%%%%%%%%%%%%%%%%%%%%%%%%%%%%%%%%%%%%%%%%%%%%%%%%%%

Due to the much larger energy separation between the $K$ and $K'$ valley for holes than for electrons, the carrier transfer is significantly slower for these. We find a  carrier generation rate at $K'$ for holes within the A-band of only 0.004$/$ps for electron-electron scattering and 0.0006$/$ps for electron-phonon scattering,

\subsection{Optical Excitation}
\label{sec_optexc}

After having used artificial occupations in Sec.\ref{sec_artexc} in order to determine carrier relaxation timescales and the importance of underlying mechanisms, we investigate here signatures of carrier relaxation under more realistic optical excitation conditions. We excite the system with a Gaussian pulse with an envelope of $E_0 exp\left(-(t-t_0)^2/\Delta_t^2\right)$, with a width $\Delta_t$ of 50$\,$fs. Since we are not concerned with details of the polarisation dynamics we do not evaluate the pertinent microscopic scattering processes but use a simple phenomenological dephasing rate $\hbar\gamma=30\,$meV. The central frequency of the pulse is 800$\,$meV above the unrenormalized bandgap. While the pulse is present, carriers created by it lead to a dynamic renormalization of the bandgap\cite{gain2}. As we have shown in Sec.\ref{sec_artexc}, these renormalizations can be comparable to the spectral width of the 50$\,$fs pulse. Thus, the detuning between central frequency of the pulse and the renormalized bandgap changes during the pulse which leads to an excitation that is spectrally broader than the one created by the artificial instantaneous excitation. 

\begin{figure}[htbp]
    \includegraphics[width=0.99\linewidth]{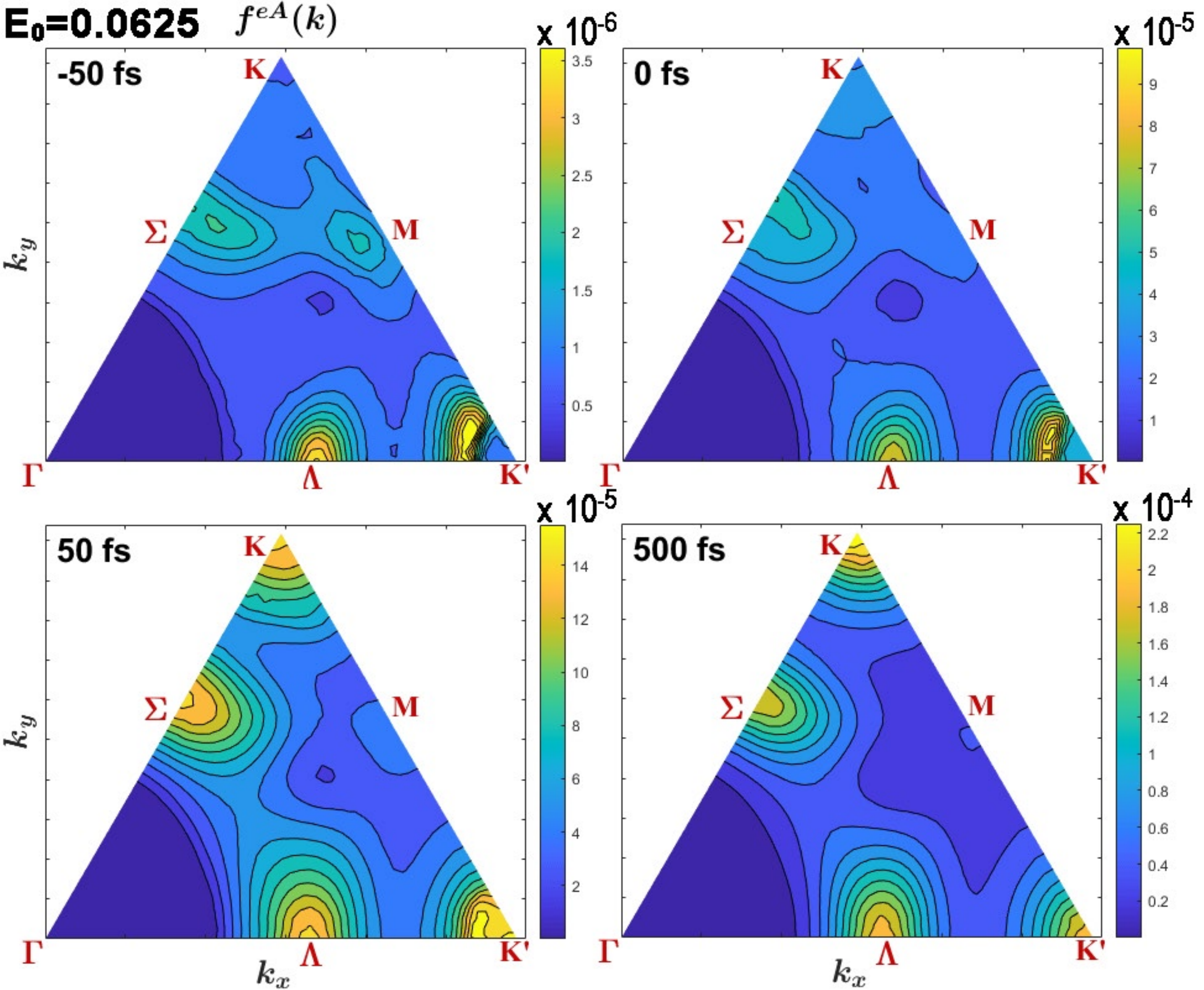}
    \includegraphics[width=0.99\linewidth]{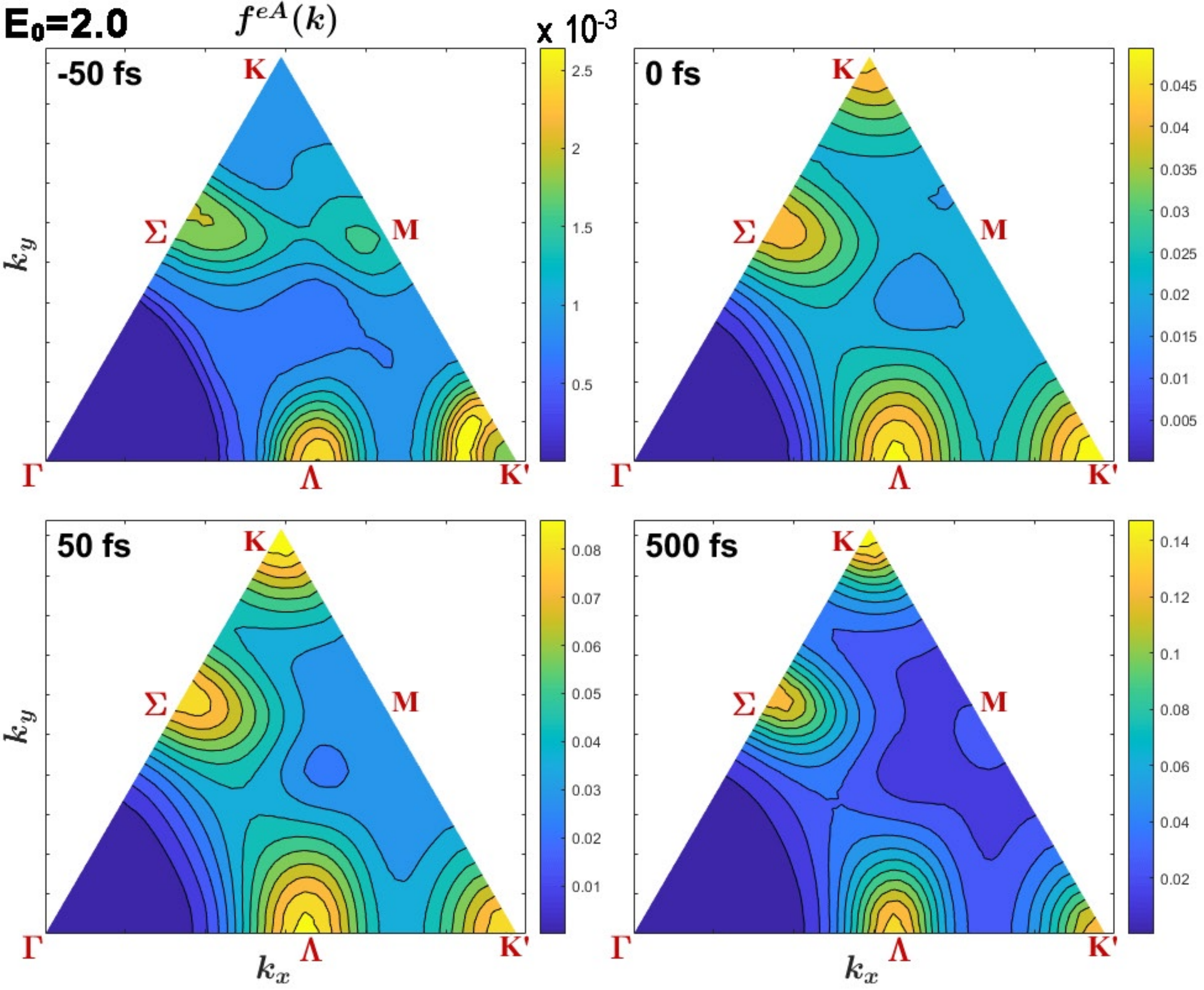}
    \caption{Occupation of the electron A-band at various times during optical excitation with a 50$\,$ fs long Gaussian pulse. The pulse maximum is at 0$\,$fs and the excitation energy is 800$\,$meV above the renormalized bandgap. Top (bottom): For an optical field $E_0=0.0625\,$MV/cm ($E_0=2.00\,$MV/cm).}
    \label{fig7}
\end{figure}

Fig.\ref{fig7} shows the occupations in the electron A-band of a monolayer of MoTe$_2$ at various times during the excitation and for two optical field strength, $E_0=0.0625\,$MV/cm and $E_0=2.00\,$MV/cm. These pulses create electron densities of about  $1\times 10^{11}/$cm$^2$ and $8\times 10^{13}/$cm$^2$, respectively.
For low excitation the carrier occupations stay below one percent at all times and smaller than 10$^{-4}$ until after the center of the pulse passed. For these low occupations electron-electron scattering is very weak. Carriers remain near the original excitation until after the pulse maximum has passed and relax into the side valleys on a ten femtosecond timescale rather than the single femtosecond scale seen for higher excitation levels in Sec.\ref{sec_artexc}. Some deviations from the strictly monotone energy dependence of the quasi Fermi distributions can still be seen 50$\,$fs after the pulse maximum at the $K'$ point. 

For the strong excitation carriers relax into hot quasi-Fermi distributions already during the pulse. At the pulse maximum carriers are already predominantly relaxed to the bottom of all valleys and the only non-thermal signature is the fact that occupations at the individual valleys are not increasing with decreasing valley energy and are, e.g. higher at $\Lambda$ and $K'$ than at $K$. The subsequent cooling down of the carriers happens on a similar timescale as for lower excitation.
 
\subsection{Energy renormalizations}
\label{sec_eneren}

\begin{figure}[htbp]
    \includegraphics[width=0.99\linewidth]{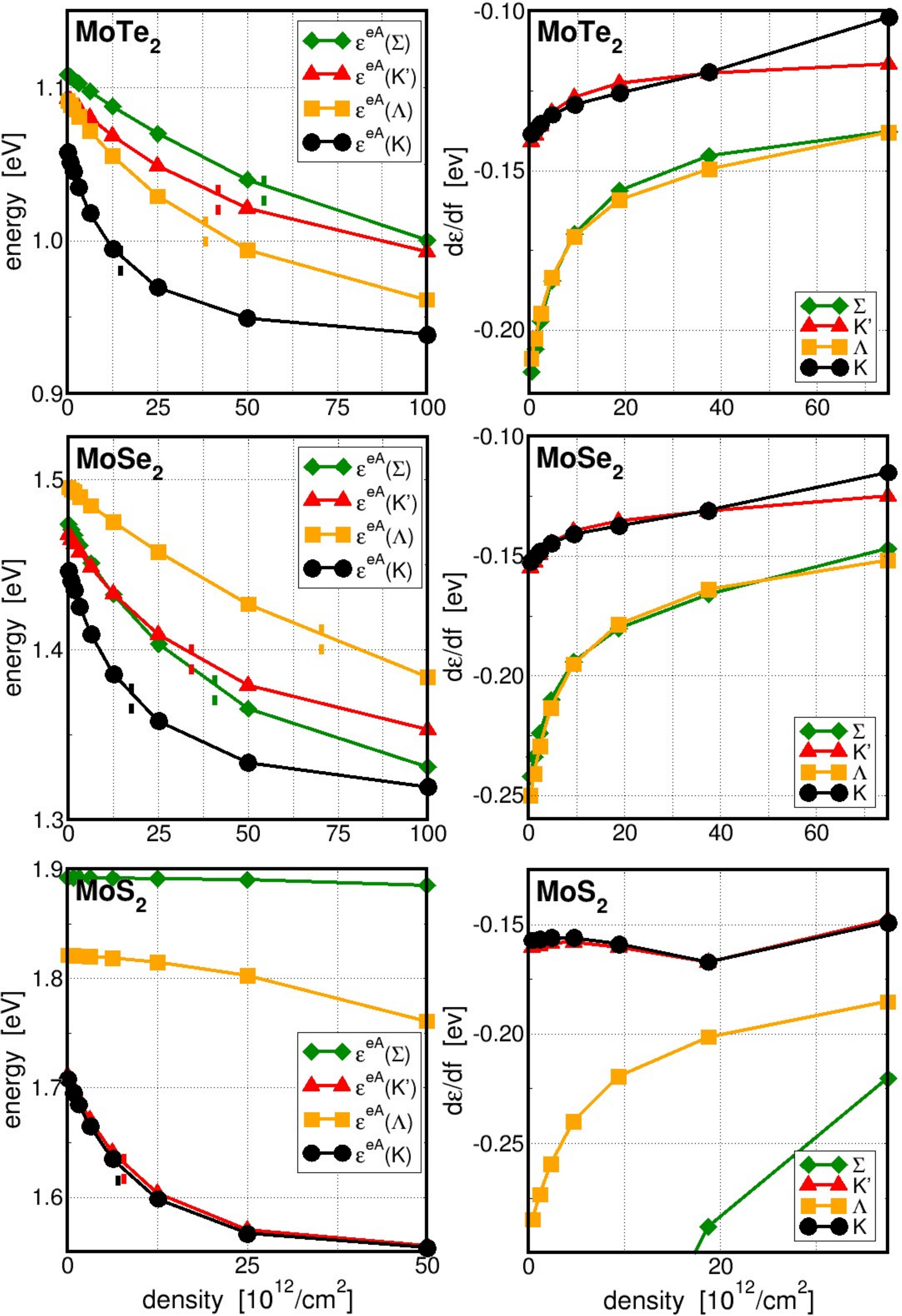}
    \caption{Left: renormalized energies of the electron A-band at critical points in the BZ as function of the electron density for monolayers of MoTe$_2$, MoSe$_2$ and MoS$_2$. Short dashed vertical lines mark the density at which the occupations are 50$\%$. Right: change of energies with increase of occupation.}
    \label{fig8}
\end{figure}

In the next step, we evaluate the density dependence of the energy renormalizations for monolayers of MoTe$_2$, MoSe$_2$ and MoS$_2$ with carriers in global thermal equilibrium  in order to check whether a transition from direct gap to indirect gap occurs. Fig.\ref{fig8} shows the computed renormalizations at the points $K$, $K' $, $\Sigma$ and $\Lambda$ for these materials. At the highest densities considered here, the occupations at the K-point of the electron A-band reach reach values above 90$\%$. For MoTe$_2$ the occupation at the highest density of $10^{14}/$cm$^2$ is about 95$\%$, for MoSe$_2$ it is about 93$\%$, and for MoS$_2$ the occupation at the highest density of $0.55\times 10^{14}/$cm$^2$ is about 97$\%$. We use these high percentages in order to test the extreme limits of the renormalizations, well knowing that realistic densities for technical applications such as lasers create electron occupations of only about 50$\%$ which is sufficient to generate inversion in the gain regime since the hole occupation at the $K$-point are usually larger due to the lack of side valleys in the hole bandstructure. To provide benchmark, we therefore mark in Fig.\ref{fig8} the densities at which the occupations reach the 50$\%$ level.

Alltogether, we never observe a transition to an indirect bandstructure for any of the materials investigated here. For all realistic densities, the lowest electron energy always remains at the $K$-point. For all materials, the change in energy with change in occupation is stronger in the $\Sigma$ and $\Lambda$ side valleys than at the $K$ and $K'$ points (see the right hand side panels of Fig.\ref{fig8}). This difference in renormalizations is caused in part by the valley dependence of the Coulomb interaction strength due to its dependence on the local wavefunction overlaps. It is also a consequence of the local bandstructure dispersion. In valleys with lower effective mass and, thus, lower density of states the occupations change stronger for a given change the in overall excitation level. However, since the side valleys are energetically significantly above the $K$-valley energies, their occupations in thermal equilibrium are low such that the local energy renormalizations are not strong enough to catch up to or even go below the $K$-point energy. Only in the hypothetical case where the energy separations in the unexcited ystems were smaller, the occupations in all valleys would be of similar order and a transition from direct to indirect would be possible.

The DFT determined energetic differences between the side- and the main-valley used in our work are considerably larger than those assumed in Ref.\cite{dirindir1} where, e.g., the valley splitting at zero density in MoS$_2$ was assumed to be less than 15$\,$meV while our calculations yield a separation of 113$\,$meV (see Table\ref{tabenee}). 
For MoSe$_2$, we find a direct gap with a minimum valley separation of about 30$\,$meV while Ref.\cite{dirindir1} assumes an indirect bandgap for that material. Our determination of a direct bandgap for MoSe$_2$ agrees with most of the literature (see e.g. Refs.\cite{mose2gap1,mose2gap2} and Refs. therein). Alltogether, the larger energy separations are the main reason why our calculations do not predict any transition to an indirect band configuration in contrast to the findings of Ref.\cite{dirindir1}. 

\section{Summary and Outlook}
\label{sec_summary}

We combine first principle DFT calculations with fully microscopic many-body models based on the semiconductor Dirac-Bloch equations in order to study the carrier dynamics and excitation induced energy renormalizations in monolayer TMDCs. Quantum Boltzmann type scattering equations for the electron-electron and electron-phonon scattering are solved taking into account the full BZ in order to resolve the detailed inter- and intra-valley relaxation processes.

For excitation high above the bandgap, we find that the carriers relax into hot quasi-Fermi distributions at the valley minima within less than 10$\,$fs. This initial relaxation is dominated by electron-electron scatterings. For the subsequent intra-valley relaxation, electron-electron and electron-phonon scatterings turn out to be of comparable importance. The subsequent cooling of the hot plasma is mediated by phonon emission on a picosecond timescale. For resonant excitation at the $K$-gap, the subsequent transfer of carriers to other valleys is found to be dominated by phonon scatterings on a single picosecond timescale.

We find that non-equilibrium signatures in the electron distribution as caused by a spectrally narrow optical excitation are only observable in the low excitation regime. Electron-electron scattering processes on the same timescale of typical femtosecond pulses hide details of the excitation at higher excitation levels.

Finally, we show that for monolayer TMDCs with a direct bandgap at zero excitation and an energetic separation between the main valley and side valleys on the order of the thermal energy or more,  like MoTe$_2$, MoSe$_2$ and MoS$_2$, the valley dependent energy renormalizations do not lead to a transition from direct to indirect bandgap. While the energy renormalization with increasing density can be stronger in the side valleys, at realistic densities it is not sufficient to shift the side-valley energetically below the $K$-point energy for any of the materials investigated here.

An investigation of further mono- and multi-layer TMDC materials and their heterostructures is required in order to see whether one of them allows for such a transition from direct to indirect due to smaller valley separations, and/or a significantly valley-dependent Coulomb strength and/ or density of states (effective mass). For example, an excitation induced  transition from an indirect to a direct gap configuration might be possible in TMDC bi-layer systems.

\section*{Data Availability}
All data that support the findings of this study are included within the article.

\begin{acknowledgments}
The authors thank the HRZ Marburg and CSC-Goethe-HLR Frankfurt for computational resources. The Tucson work was supported by the Air Force Office of Scientific Research under award numbers FA9550-17-1-0246 and FA9550-21-1-0463.
\end{acknowledgments} 

\section*{ORCID iDs}

J{\"o}rg Hader  https://orcid.org/0000-0003-1760-3652\\
Josefine Neuhaus  https://orcid.org/0000-0003-1440-1588\\
Jerome V. Moloney  https://orcid.org/0000-0001-8866-0326\\
Stephan W. Koch  https://orcid.org/0000-0001-5473-0170

\end{document}